\documentclass[aps,prd,preprint,amsmath,groupedaddress,showpacs]{revtex4}

\usepackage{graphicx}
\usepackage{dcolumn}
\usepackage{bm}
\usepackage{color}
\usepackage{hyperref}
\usepackage{verbatim}

\begin{document}

\title{Higgs boson $ZZ$ couplings in Higgs-strahlung at the ILC}

\author{Taras Zagoskin}
\email{taras.zagoskin@gmail.com}
\author{Alexander Korchin}
\email{korchin@kipt.kharkov.ua}
\affiliation{NSC ``Kharkov Institute of Physics and Technology'', 61108 Kharkov, Ukraine \\
V.N.~Karazin Kharkov National University, 61022 Kharkov, Ukraine}

\date{\today}

\begin{abstract}
We derive the fully differential cross section of the Higgs-strahlung process $f \bar{f} \to Z \to Z (\to f_Z \bar{f}_Z) X (\to f_X \bar{f}_X)$, where $f$, $f_Z$, and $f_X$ are arbitrary fermions and $X$ is a spin-zero particle with arbitrary couplings to $Z$ bosons and fermions. This process with $f = e$ and $X = h$ ($h$ denotes the Higgs boson) is planned to be measured at the ILC to put constraints on the couplings $g_1$, $g_2$, $g_3$ of the Higgs boson to $Z$ bosons. Using the obtained fully differential cross section, we define observables measurement of which yields the tightest constraints on the couplings. Explicit dependences of these observables on $g_i$ are derived. 
\end{abstract}
\pacs{12.15.Ji, 13.66.Fg, 14.80.Bn.}

\maketitle

\section{Introduction}
\label{Section: Introduction}

Since the discovery \cite{Aad:2012, Chatrchyan:2012} of the Higgs boson (denoted by $h$ in this paper) in 2012, it has been important to measure its couplings and CP properties. These measurements can prove or disprove the Higgs mechanism \cite{the_Higgs_mechanism} and can let us describe processes involving the Higgs boson more precisely. 

The couplings of the Higgs boson are measured at the LHC by the CMS and ATLAS collaborations (see, for example, \cite{the_hZZ_couplings_CMS, the_hZZ_hWW_and_hgg_couplings_ATLAS, the_Higgs_boson_couplings_CMS_and_ATLAS}). These couplings are also planned\cite{the_Linear_Collider_Collaboration_strategy_on_measuring_the_Higgs_boson_properties} to be measured at the ILC. The latter measurements are expected to have lower backgrounds and to yield stricter constraints on a number of the Higgs boson parameters. Many papers have addressed the prospects of measurements at the ILC \cite{the_prospects_of_measurements_at_the_ILC_1, the_prospects_of_measurements_at_the_ILC_2, the_prospects_of_measurements_at_the_ILC_3, the_prospects_of_measurements_at_the_ILC_4, the_prospects_of_measurements_at_the_ILC_5, the_prospects_of_measurements_at_the_ILC_6, the_prospects_of_measurements_at_the_ILC_7, the_prospects_of_measurements_at_the_ILC_8, the_prospects_of_measurements_at_the_ILC_9, the_prospects_of_measurements_at_the_ILC_10}. 

At the ILC, the Higgs couplings to a pair of $Z$ bosons ($hZZ$ couplings) will be measured in the Higgs-strahlung process $e^- e^+ \to Z \to Z (\to f_Z \bar{f}_Z) h (\to f_h \bar{f}_h)$, where $f_Z = e, \mu, u, d, s, c, b$ and $f_h = \tau, b$. A lof of papers (see, for example, Refs.~\cite{paper_1_on_ee-->Z-->Z(-->ff)h, paper_2_on_ee-->Z-->Z(-->ff)h, paper_3_on_ee-->Z-->Z(-->ff)h, paper_4_on_ee-->Z-->Z(-->ff)h}) concern the process $e^- e^+ \to Z \to Z (\to f_Z \bar{f}_Z) h$, not considering a decay of the Higgs boson. Consideration of such a decay allows for the fact that the Higgs boson is off-shell, thereby allowing us to obtain more precise total and differential cross sections and observables of the Higgs-strahlung. Moreover, the Higgs-strahlung with a decay of the Higgs boson is studied in Ref.~\cite{the_Higgs-strahlung_without_the_positron_beam_polarization} (see Eq.~(A2) there). We generalize Eq.~(A2) to the case of an arbitrary polarization of the positron beam.

In Section~\ref{Section: Fully differential cross section} we derive the fully differential cross section of this process for a spin-zero Higgs boson, accounting for the beyond the Standard Model (SM) $hZZ$ and $hff$ couplings and for arbitrary electron and positron polarizations. In Section~\ref{Section: Optimal observables} we use the obtained formula and Refs.~\cite{optimal_observables_---_a_one-dimensional_case, optimal_observables_---_a_multidimensional_case} to define observables yielding the tightest constraints on the $hZZ$ couplings. The dependence of these observables on the couplings is derived and analyzed. 

\section{Fully differential cross section}
\label{Section: Fully differential cross section}

\begin{figure}
\includegraphics[scale=0.7]{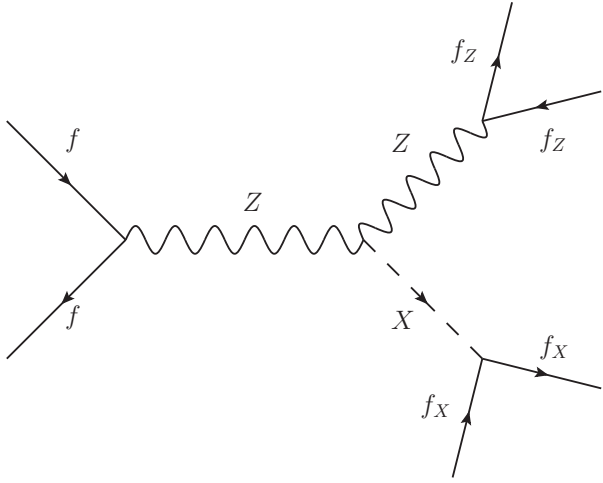}
\caption{A Feynman diagram of the Higgs-strahlung.}
\label{A Feynman diagram of the Higgs-strahlung}
\end{figure}

We consider the process
\begin{align} 
\label{the Higgs-strahlung}
f \bar{f} \to Z \to Z X \to f_Z \bar{f}_Z f_X \bar{f}_X
\end{align}
(see Fig.~\ref{A Feynman diagram of the Higgs-strahlung}), where $f$, $f_Z$, and $f_X$ are some fermions, $f_Z \neq f_X$, and $X$ is a particle with zero spin, arbitrary couplings to a pair of $Z$ bosons and arbitrary couplings to a fermion-antifermion pair. This scattering is going to be measured at the ILC in the case $f = e$, $X = h$, $f_Z = e, \mu, u, d, s, c, b$, and $f_h = \tau, b$. 

Due to the energy-momentum conservation in process (\ref{the Higgs-strahlung}), 
\begin{align} 
\label{the masses of the two Z bosons and the X boson}
a_1 = s, \qquad a_2 \in (4 m_{f_Z}^2, (\sqrt{a_1} - \sqrt{a_X})^2), \qquad a_X \in (4 m_{f_X}^2, (\sqrt{a_1} - 2 m_{f_Z})^2),
\end{align}
where $a_1$, $a_2$, and $a_X$ are the squared invariant masses of the $Z$ boson (called $Z_1$) produced by the fermion-antifermion pair $f \bar{f}$, of the $Z$ boson (called $Z_2$) produced together with the boson $X$, and of the boson $X$ itself, respectively, $s$ is the squared invariant energy of $f \bar{f}$, $m_{f_Z}$ and $m_{f_X}$ are the masses of the fermions $f_Z$ and $f_X$ respectively. 

The amplitude $A (\lambda_1, \lambda_2)$ of the transition $Z \to ZX$ is analogous to the $HVV$ amplitude in Ref.~\cite{the_hZZ_couplings_CMS} (see Eq.~(1) there) and to the $XZZ$ amplitude in Ref.~\cite{the_XZZ_amplitude_1} (Eq.~(7)) and in \cite{the_XZZ_amplitude_2} (Eq.~(3)): 
\begin{align}
\label{definitions_gi}
A (\lambda_1, \lambda_2) = & 2 \sqrt{\sqrt{2} G_F} m_Z^2 \Biggl (\left (g_1 (a_1, a_2) - \frac{a_1 + a_2 - a_X}{m_Z^2} g_2 (a_1, a_2) \right) (e_1 \cdot e_2^*) & \notag \\
& + 2 \frac {g_2 (a_1, a_2)} {m_Z^2} (e_1 \cdot q_2) (e_2^* \cdot q_1) + 2 \frac {g_3 (a_1, a_2)} {m_Z^2} \varepsilon_{\mu \nu \rho \sigma} q_1^{\mu} q_2^{\nu} e_1^{\rho} (e_2^{\sigma})^* \Biggr), &
\end{align}
where $\lambda_j$, $e_j$, and $q_j$ are the helicity, polarization 4-vector, and 4-momentum of the boson $Z_j$ respectively ($j = 1, 2$), $G_F$ is the Fermi constant, $m_Z$ is the pole mass of the $Z$ boson, $g_1 (a_1, a_2)$, $g_2 (a_1, a_2)$, and $g_3 (a_1, a_2)$ are some complex-valued functions on $a_1$ and $a_2$ --- we call these functions $XZZ$ couplings, $\varepsilon_{\mu \nu \rho \sigma}$ is the Levi-Civita symbol ($\varepsilon_{0 1 2 3} = 1$). In Ref.~\cite{the_hZZ_couplings_CMS} the $XZZ$ couplings are denoted as $a_1$, $a_2$, and $a_3$ --- we denote them as $g_1$, $g_2$, and $g_3$ respectively to avoid confusion. 

At the tree level, the $XZZ$ couplings are connected with the CP parity of the boson $X$, as shown in Table~\ref{The CP parity of X in the case of various sets of values of the XZZ couplings}. In the SM $g_1 = 1$ and $g_2 = g_3 = 0$. 
%%%%%%%%%%% Table %%%%%%%%%%%%%
\begin{table}[h]
\caption{The $CP$ parity of the particle $X$ for various values of $g_1$, $g_2$, and $g_3$.}
\label{The CP parity of X in the case of various sets of values of the XZZ couplings}
\begin{tabular}[t]{c c c | c}
\hline
$g_1$ & $g_2$ & $g_3$ & $CP_X$ \\
\hline
{\rm any} & {\rm any} & 0 & 1\\ 
\hline
0 & 0 & $\neq$ 0 & $-1$ \\
\hline
$\neq$ 0 & {\rm any} & $\neq$ 0 & {\rm indefinite} \\
{\rm any} & $\neq 0$ & $\neq 0$ & \\
\hline
\end{tabular}
\end{table}
%%%%%%%%%%%%%%%%%%%%%%%%%%%

In the $Z_1$ rest frame, or the center-of-mass frame of process (\ref{the Higgs-strahlung}), the polarization vectors read:
\begin{align}
\label{the polarization 4-vector of Z_1}
& e_1^{\mu}|_{\lambda_1 = 0} = (0, 0, 0, 1), \qquad e_1^{\mu}|_{\lambda_1 = \pm 1} = (0, \mp \frac{1}{\sqrt{2}}, \frac{i}{\sqrt{2}}, 0),  & \\
\label{the polarization 4-vector of Z_2}
& e_2^{\mu}|_{\lambda_2 = 0} = (\frac{|\mathbf{q_2}|}{\sqrt{a_2}}, 0, 0, - \frac{\sqrt{|\mathbf{q_2}|^2 + a_2}}{\sqrt{a_2}}), \qquad e_2^{\mu}|_{\lambda_2 = \pm 1} = (0, \pm \frac{1}{\sqrt{2}}, \frac{i}{\sqrt{2}}, 0). &
\end{align}

Using Eqs.~(\ref{definitions_gi}), (\ref{the polarization 4-vector of Z_1}) and (\ref{the polarization 4-vector of Z_2}), we derive that
\begin{align}
& A (-1, 1) = - g_1 - \frac{a_X - a_1 - a_2}{m_Z^2} g_2 + 2 i \frac{\sqrt{a_1} |\mathbf{q_2}|}{m_Z^2} g_3, & \notag \\ 
& A (1, -1) = - g_1 - \frac{a_X - a_1 - a_2}{m_Z^2} g_2 - 2 i \frac{\sqrt{a_1} |\mathbf{q_2}|}{m_Z^2} g_3, & \notag \\ 
& A (0, 0) = (g_1 + \frac{a_X - a_1 - a_2}{m_Z^2} g_2) \frac{\sqrt{|\mathbf{q_2}|^2 + a_2}}{\sqrt{a_2}} + 2 \frac{\sqrt{a_1} |\mathbf{q_2}|^2}{m_Z^2 \sqrt{a_2}} g_2, & \notag \\ 
& A (\lambda_1, \lambda_2) = 0 \quad  {\rm for}  \quad \lambda_1 \neq - \lambda_2. &
\end{align}
 
The amplitude of the decay $X \to f_X \bar{f}_X$ is \cite{the_amplitude_of_the_X_to_f_antif_transition}
\begin{align} 
\label{the amplitude of the X to f antif transition}
A_{X \to f_X \bar{f}_X} (\lambda_{f_X}, \lambda_{\bar{f}_X}) = - \sqrt{\sqrt{2} G_F} m_{f_X} \bar{u}_{f_X} (\rho_1 + \rho_2 \gamma^5) v_{\bar{f}_X}, 
\end{align}
where $u$ and $v$ are the Dirac spinors, $\lambda_{f_X}$ ($\lambda_{\bar{f}_X}$) is the helicity of the fermion (antifermion), $\rho_1$ and $\rho_2$ are some complex numbers which we call the $Xff$ couplings.  In the SM $\rho_1 = 1$ and $\rho_2 = 0$. 

Calculation of $A_{X \to f_X \bar{f}_X} (\lambda_{f_X}, \lambda_{\bar{f}_X})$ in the $X$ rest frame yields 
\begin{align} 
\label{the amplitude of the X to f antif transition}
& A_{X \to f_X \bar{f}_X} (- \frac{1}{2}, - \frac{1}{2}) = \sqrt{\sqrt{2} G_F} m_{f_X} (\rho_1 + \rho_2) \sqrt{a_X}, & \notag \\
& A_{X \to f_X \bar{f}_X} (\frac{1}{2}, \frac{1}{2}) = \sqrt{\sqrt{2} G_F} m_{f_X} (\rho_1 - \rho_2) \sqrt{a_X}, & \notag \\
& A_{X \to f_X \bar{f}_X} (- \frac{1}{2}, \frac{1}{2}) = A_{X \to f_X \bar{f}_X} (\frac{1}{2}, - \frac{1}{2}) = 0. &
\end{align}

\begin{figure}
\includegraphics[scale=0.7]{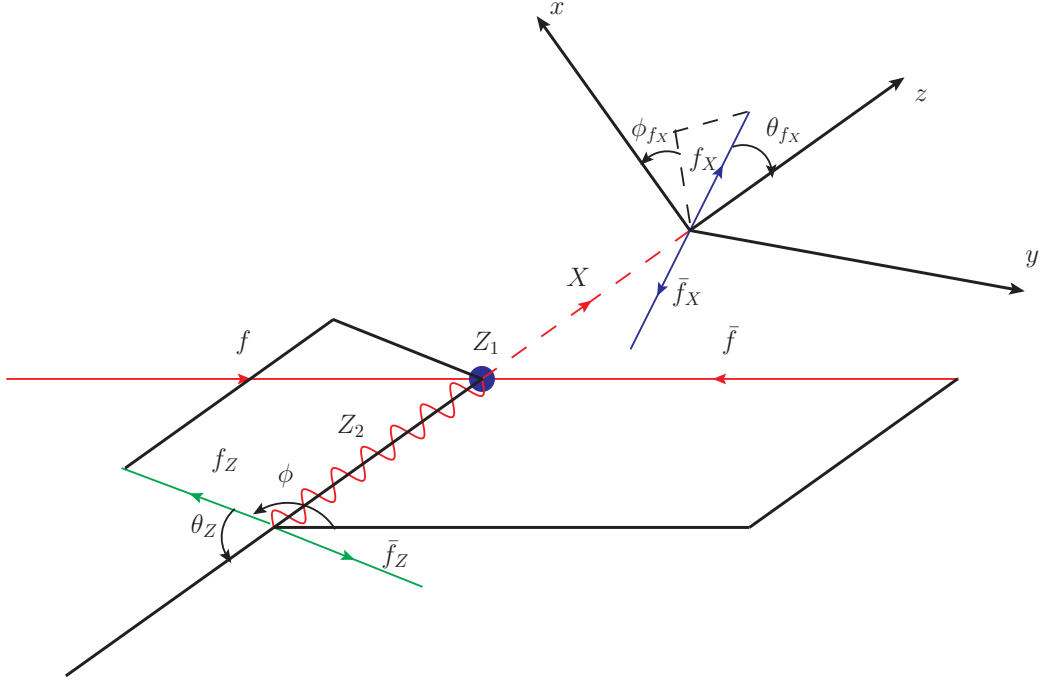}
\caption{Kinematics of process (\ref{the Higgs-strahlung}). The momenta of $f$, $\bar{f}$, $Z_1$, $X$, and $Z_2$ are shown in the $f \bar{f}$ rest frame, the momenta of $f_Z$ and $\bar{f}_Z$ are displayed in the $Z_2$ rest frame, the momenta of $f_X$ and $\bar{f}_X$ are described in the $X$ rest frame. The $z$ axis is co-directional with the $X$ momentum while the $x$ and $y$ axes are arbitrary axes forming a right-handed system with the $z$ axis.}
\label{kinematics of process (1)}
\end{figure}

Using the helicity formalism and neglecting $m_f$, $m_{f_Z}$, and $m_{f_X}$ everywhere save the $m_{f_X}$ factor in Eq.~(\ref{the amplitude of the X to f antif transition}), we derive the fully differential cross section of process (\ref{the Higgs-strahlung}):
\begin{align} 
\label{the fully differential cross section}
& \frac{d^7 \sigma}{d a_2 d a_X d \theta_Z d \theta_{f_Z} d \phi d \theta_{f_X} d \phi_{f_X}} = \frac{G_F^4 m_Z^8 m_{f_X}^2}{(4 \pi)^7} \cdot \frac{a_2 a_X \lambda^{1/2} (s, a_2, a_X)}{s D (s) D (a_2) D_X (a_X)} (a_f^2 + v_f^2) (a_{f_Z}^2 + v_{f_Z}^2) & \notag \\
& \times (|\rho_1|^2 + |\rho_2|^2) \sin \theta_Z \sin \theta_{f_Z} \sin \theta_{f_X} \Biggl [ (|A_{\parallel}|^2 + |A_{\perp}|^2) \Bigl (P_1 (1 + \cos^2 \theta_Z) (1 + \cos^2 \theta_{f_Z}) & \notag \\
& + 4 P_2 A_{f_Z} \cos \theta_Z \cos \theta_{f_Z} \Bigr) + {\rm Re} (A_{\parallel}^* A_{\perp}) \cdot 4 \Bigl (P_1 A_{f_Z} \cos \theta_{f_Z} (1 + \cos^2 \theta_Z) & \notag \\
& + P_2 \cos \theta_Z (1 + \cos^2 \theta_{f_Z}) \Bigr) + |A_0|^2 \cdot 4 P_1 \sin^2 \theta_Z \sin^2 \theta_{f_Z} - 4 \sqrt{2} \sin \theta_Z \sin \theta_{f_Z} & \notag \\
& \times \Bigl ( ({\rm Re} (A_0^* A_{\parallel}) \cos \phi + {\rm Im} (A_0^* A_{\perp}) \sin \phi) (P_2 A_{f_Z} + P_1 \cos \theta_Z \cos \theta_{f_Z}) & \notag \\
& + ({\rm Im} (A_0^* A_{\parallel}) \sin \phi + {\rm Re} (A_0^* A_{\perp}) \cos \phi) (P_1 A_{f_Z} \cos \theta_Z + P_2 \cos \theta_{f_Z}) \Bigr) + P_1 \sin^2 \theta_Z \sin^2 \theta_{f_Z} & \notag \\
& \times \Bigl ( (|A_{\parallel}|^2 - |A_{\perp}|^2) \cos 2 \phi + {\rm Im} (A_{\parallel}^* A_{\perp}) \cdot 2 \sin 2 \phi \Bigr) \Biggr], &
\end{align}
where (see Fig.~\ref{kinematics of process (1)})
\begin{itemize}
\item $\theta_Z$ is the angle between the momentum $\mathbf{p}_f$ of the fermion $f$ in the $f \bar{f}$ rest frame and the momentum $\mathbf{q}_2$ of the $Z$ boson $Z_2$ in the same frame;
\item $\theta_{f_Z}$ is the angle between $\mathbf{q}_2$ and the $f_Z$ momentum $\mathbf{p}_{f_Z}$ in the $Z_2$ rest frame;
\item $\phi$ is the azimuthal angle between the plane spanned by the vectors $\mathbf{p}_f$ and $\mathbf{q}_2$ and the plane spanned by $\mathbf{q}_2$ and $\mathbf{p}_{f_Z}$;
\item $\theta_{f_X}$ is the angle between the $X$ momentum in the $f \bar{f}$ rest frame and the $f_X$ momentum in the $X$ rest frame;
\item $\phi_{f_X}$ is the azimuthal angle of the fermion $f_X$;
\item $\lambda (x, y, z) \equiv x^2 + y^2 + z^2 - 2 (xy + xz + yz)$;
\item $D (x) \equiv (x - m_Z^2)^2 + (m_Z \Gamma_Z)^2$, \ $D_X (x) \equiv (x - m_X^2)^2 + (m_X \Gamma_X)^2$;
\item $\Gamma_Z$ ($\Gamma_X$) is the total width of the $Z$ ($X$) boson;
\item $m_X$ is the pole mass of the $X$ boson;
\item $a_f$ is the projection of the weak isospin of a fermion $f$, $v_f \equiv a_f - 2 \frac{q_f}{e} \sin^2 \theta_W$, $q_f$ is the electric charge of the fermion $f$, $e$ is the electric charge of the positron, $\theta_W$ is the weak mixing angle, $A_f \equiv \frac{2 a_f v_f}{a_f^2 + v_f^2}$;
\item $A_0 \equiv \frac{s + a_2 - a_X}{2 \sqrt{s a_2}} g_1 - \frac{2 \sqrt{s a_2}}{m_Z^2} g_2$, 
$A_{\parallel} \equiv - \frac{A (-1, 1) + A (1, -1)}{\sqrt{2}} = \sqrt{2} \left (g_1 - \frac{s + a_2 - a_X}{m_Z^2} g_2 \right)$,
$A_{\perp} \equiv - \frac{A (-1, 1) - A (1, -1)}{\sqrt{2}} = i \sqrt{2} \frac{\lambda^{1/2} (s, a_2, a_X)}{m_Z^2} g_3$; 
\item $P_1 \equiv 1 - P_f P_{\bar{f}} - (P_f - P_{\bar{f}}) A_f$, \  $P_2 \equiv (1 - P_f P_{\bar{f}}) A_f - P_f + P_{\bar{f}}$, 
where  $P_f$ ($P_{\bar{f}}$) is the fermion (antifermion) beam longitudinal polarization.
\end{itemize}

According to Eq.~(\ref{the fully differential cross section}), all the possible directions of the $f_X$ momentum in the $X$ rest frame are equiprobable. Having measured an observable describing Higgs-strahlung (\ref{the Higgs-strahlung}), one will be able to use Eq.~(\ref{the fully differential cross section}) and to put some constraints on the functions $A_0$, $A_{\parallel}$, and $A_{\perp}$, thus getting possible intervals for the $XZZ$ couplings $g_1$, $g_2$, and $g_3$. 

The differential cross section of the process $e^- e^+ \to f \bar{f} h$ is obtained in Ref.~\cite{the_Higgs_strahlung_fully_differential_cross_section_without_consideration_of_the_Higgs_boson_decay}. However, considering process (\ref{the Higgs-strahlung}), we take into account that the boson $X$ decays, which happens in reality. Moreover, integration of Eq.~(\ref{the fully differential cross section}) yields more precise total and differential cross sections and observables than those which can be derived by integration of the fully differential cross section of $e^- e^+ \to f \bar{f} h$. The reason is that we can integrate (\ref{the fully differential cross section}) with respect to $a_X$ without the narrow-width approximation and with any desired accuracy.

Integration of Eq.~(\ref{the fully differential cross section}) with the narrow-width approximation for both $Z$ and $X$ boson yields the total cross section of the Higgs-strahlung:
\begin{align} 
\label{the total cross section}
\sigma = & \frac{G_F^4 m_Z^9 m_X m_{f_X}^2}{288 \pi^3 \Gamma_Z \Gamma_X} \cdot \frac{\lambda^{1/2} (s, m_Z^2, m_X^2)}{s D (s)} (a_f^2 + v_f^2) (a_{f_Z}^2 + v_{f_Z}^2) (|\rho_1|^2 + |\rho_2|^2) P_1 & \notag \\
& \times \sum_{p = 0, \parallel,\perp} |A_p|^2 \Bigl |_{a_2 = m_Z^2, \, a_X = m_X^2}. &
\end{align}

Since $\sigma \sim P_1$, the total cross section has its largest value at $P_f = - 1$ and $P_{\bar{f}} = 1$ if $A_f > 0$.

\section{Optimal observables}
\label{Section: Optimal observables}

In this section we demonstrate how differential cross section (\ref{the fully differential cross section}) can be used to determine the couplings $g_1, g_2, g_3$ after kinematic information on a number of events (\ref{the Higgs-strahlung}) is obtained. For this purpose, the method of Refs.~\cite{optimal_observables_---_a_one-dimensional_case, optimal_observables_---_a_multidimensional_case} is applied.
According to~\cite{optimal_observables_---_a_one-dimensional_case, optimal_observables_---_a_multidimensional_case}, if the fully differential cross section of a process can be presented in the form 
\begin{align}
\label{the fully differential cross section of a process with small parameters}
\frac{d \sigma}{d k} = S_0 (k) + \sum \limits_{i = 1}^{n} h_i S_i (k),
\end{align}
where $k$ denotes all the kinematic variables of this process, $S_0 (k), S_1 (k), \ldots, S_n (k)$ are functions of $k$, \ $h_i$ ($i = 1, \ldots, n$) are real-valued small (i.e. $h_i \to 0$) dimensionless parameters independent of $k$, then measurement of an observable
\begin{align}
\label{a_definition_of_an_optimal_observable}
O_i \equiv \frac{1}{\sigma} \int d k \frac{d \sigma}{d k} \cdot \frac{S_i (k)}{S_0 (k)}
\end{align}
yields the tightest constraint on the parameter $h_i$. ``The tightest constraint on a parameter'' means an experimental value of the parameter with the least standard deviation which is possible for measured data for the studied process. In (\ref{a_definition_of_an_optimal_observable}) $\sigma$ is the total cross section of the process:
\begin{align}
\label{a_definition_of_the_total_cross_section_of_a_process}
\sigma \equiv \int d k \frac{d \sigma}{d k}.
\end{align}
The integration in Eqs.~(\ref{a_definition_of_an_optimal_observable}) and (\ref{a_definition_of_the_total_cross_section_of_a_process}) is performed over all the possible values of the variables $k$.  In Ref.~\cite{optimal_observables_---_a_multidimensional_case} observables (\ref{a_definition_of_an_optimal_observable}) are called optimal. 

In our case, differential cross section (\ref{the fully differential cross section}) can be rewritten as 
\begin{align}
\label{the fully differential cross section of the Higgs-strahlung}
\frac{d \sigma}{d k} = & (|\rho_1|^2 + |\rho_2|^2) \Bigr [f_{SM} (k) + (|g_1|^2 - 1) f_{1} (k) + |g_2|^2 f_2 (k)+ |g_3|^2 f_3 (k) & \notag \\
& + {\rm Re} (g_1^* g_2) f_4 (k) + {\rm Re} (g_1^* g_3) f_5 (k) + {\rm Re} (g_2^* g_3) f_6 (k) & \notag \\
& + {\rm Im} (g_1^* g_2) f_7 (k) + {\rm Im} (g_1^* g_3) f_8 (k) + {\rm Im} (g_2^* g_3) f_9 (k) \Bigl], &
\end{align}
where $d k \equiv d a_2 d a_X d \theta_Z d \theta_{f_Z} d \phi d \theta_{f_X} d \phi_{f_X}$ and $f_{SM} (k), f_1(k) =f_{SM}(k), f_2 (k), f_3 (k), \ldots, f_9 (k)$ are some particular functions of the invariant masses squared $a_2$, $a_X$, 
of the angular variables $\theta_Z$, $\theta_{f_Z}$, $\phi$, $\theta_{f_X}$, of the polarizations $P_f$, $P_{\bar{f}}$, and of the squared invariant energy $s$. 

In the SM 
\begin{align}
\frac{d \sigma}{d k} = (|\rho_1|^2 + |\rho_2|^2) f_{SM} (k). 
\end{align}
We suppose that the couplings $g_1$, $g_2$, and $g_3$ in (\ref{definitions_gi}) are close to their SM values and do not depend on $a_1$ or $a_2$, i.e., $g_1 (a_1, a_2) \approx 1$, $g_2 (a_1, a_2) \approx 0$, $g_3 (a_1, a_2) \approx 0$. Then Eq.~(\ref{the fully differential cross section of the Higgs-strahlung}) has the form of Eq.~(\ref{the fully differential cross section of a process with small parameters}) with $n = 9$ and 
\begin{align}
& S_0 (k) = S_1 (k) = (|\rho_1|^2 + |\rho_2|^2) f_{SM} (k), & \notag \\
& S_2 (k) = (|\rho_1|^2 + |\rho_2|^2) \, f_2 (k), & \notag \\
& \ldots & \notag \\
& S_9 (k) = (|\rho_1|^2 + |\rho_2|^2) \, f_9 (k), & \notag \\
& h_1 = |g_1|^2 - 1, \; h_2 = |g_2|^2, \; h_3 = |g_3|^2, & \notag \\
& h_4 = {\rm Re} (g_1^* g_2), \; \ldots , \; h_9 = {\rm Im} (g_2^* g_3). & 
\end{align}

Therefore, the observable $O_1$ for the Higgs-strahlung is 
\begin{align}
O_1 \equiv \frac{1}{\sigma} \int d k \frac{d \sigma}{d k} = 1. 
\end{align}
Thus this observable provides no information on the quantity $|g_1|^2 - 1$, i.e. no constraint on $|g_1|$. If we measure the fully differential distribution $\frac{1}{\sigma} \frac{d \sigma}{d k}$ of the Higgs-strahlung, we will gain no constraint on $|g_1|$ as well. The reason is that according to Eq.~(\ref{the fully differential cross section of the Higgs-strahlung}), the dependence of $\frac{1}{\sigma} \frac{d \sigma}{d k}$ on the couplings $g_1$, $g_2$, and $g_3$ reduces only to a dependence on the fractions $\frac{g_2}{g_1}$ and $\frac{g_3}{g_1}$. Thus, measurement of the fully differential distribution yields constraints only on the latter fractions. To constrain $|g_1|$, one can measure $\sigma$. 

We calculate numerically the observables 
\begin{align}
\label{optimal observables O2, O3, ..., O9}
O_i \equiv \frac{1}{\sigma} \int d k \frac{d \sigma}{d k} \frac{f_i (k)}{f_{SM} (k)}, \qquad i = 2, 3, \ldots, 9
\end{align}
for process (\ref{the Higgs-strahlung}) with $f = e$, using the values listed in Table~\ref{The parameters used to calculate optimal observables for the Higgs-strahlung}. In this table we show the $\sqrt{s}$, $P_{e^-}$, and $P_{e^+}$ values planned for the first stage of the ILC (see~\cite{the_Linear_Collider_Collaboration_strategy_on_measuring_the_Higgs_boson_properties}). 

\begin{table}[h]
\caption{The parameters used to calculate optimal observables (\ref{optimal observables O2, O3, ..., O9}) for the Higgs-strahlung. These parameters are taken from Refs.~\cite{the_Linear_Collider_Collaboration_strategy_on_measuring_the_Higgs_boson_properties, the_2016_review_of_particle_physics_by_the_Particle_Data_Group}. For $\Gamma_h$ we use its SM value.}
\label{The parameters used to calculate optimal observables for the Higgs-strahlung}
\begin{tabular}{l}
\hline
$P_{e^-} = - 0.8$, \ $P_{e^+} = 0.3$, \ $\sqrt{s} = 250$ GeV\\
$A_e = 0.1515$ \\
$m_h = 125.09$ GeV, \ $\Gamma_h = 4.15$ MeV \\
$m_Z = 91.1876$ GeV, \ $\Gamma_Z = 2.4952$ GeV \\
\hline
\end{tabular}
\end{table}

Observables (\ref{optimal observables O2, O3, ..., O9}) calculated for process (\ref{the Higgs-strahlung}) with $f = f_Z = e$ and $f_X = e, \mu, \tau, u, d, s, c, b$ are
\begin{align}
\label{explicit expressions for the optimal observables}
O_2 = & \, G^{-1}  \Bigl( {34.1 + 1281.8 \, |\frac{g_2}{g_1}|^2 + 311.2 \, |\frac{g_3}{g_1}|^2 - 395.4 \, {\rm Re} (\frac{g_2}{g_1})} \Bigr), & \notag \\
O_3 = & \, G^{-1}  \Bigl({8.0 + 311.2 \, |\frac{g_2}{g_1}|^2 + 82.7 \, |\frac{g_3}{g_1}|^2 - 94.0 \, {\rm Re} (\frac{g_2}{g_1})}\Bigr), & \notag \\
O_4 = & \,G^{-1}  \Bigl({- 11.5 - 395.4 \, |\frac{g_2}{g_1}|^2 - 94.0 |\frac{g_3}{g_1}|^2 + 132.9 \, {\rm Re} (\frac{g_2}{g_1})}\Bigr), & \notag \\
O_5 = & \, G^{-1}  \Bigl({1.1 \, {\rm Re} (\frac{g_3}{g_1}) - 6.6 \, {\rm Re} ((\frac{g_2}{g_1})^* \frac{g_3}{g_1})}\Bigr), & \notag \\
O_6 = & \, G^{-1}  \Bigl({- 6.6 \, {\rm Re} (\frac{g_3}{g_1}) + 40.1 \, {\rm Re} ((\frac{g_2}{g_1})^* \frac{g_3}{g_1})}\Bigr), & \notag \\
O_7 = & \,G^{-1} \cdot {0.24 \, {\rm Im} (\frac{g_2}{g_1})}, & \notag \\
O_8 = & \,G^{-1}  \Bigl({13.6 \, {\rm Im} (\frac{g_3}{g_1}) - 88.7 \, {\rm Im} ((\frac{g_2}{g_1})^* \frac{g_3}{g_1})}\Bigr), & \notag \\
O_9 = & \, G^{-1}  \Bigl({- 88.7 \, {\rm Im} (\frac{g_3}{g_1}) + 583.4 \, {\rm Im} ((\frac{g_2}{g_1})^* \frac{g_3}{g_1})}\Bigr), &
\end{align}
where 
\begin{equation}
\label{expression for G}
G \equiv 1 + 34.1 \left |\frac{g_2}{g_1} \right |^2 + 8.0 \left |\frac{g_3}{g_1} \right |^2 - 11.5 \, {\rm Re} \left( \frac{g_2}{g_1} \right).
\end{equation}

Observables (\ref{explicit expressions for the optimal observables}) depend only on the ratios $\frac{g_2}{g_1}$ and $\frac{g_3}{g_1}$. Thus their measurement gives constraints only on the real quantities ${\rm Re} (\frac{g_2}{g_1})$, ${\rm Im} (\frac{g_2}{g_1})$, ${\rm Re} (\frac{g_3}{g_1})$, and ${\rm Im} (\frac{g_3}{g_1})$. 

If the couplings $g_1$, $g_2$, and $g_3$ are real, the Lagrangian describing the interaction of the Higgs boson with two $Z$ bosons is Hermitian. Considering only real values of $\frac{g_2}{g_1}$ and $\frac{g_3}{g_1}$, we present plots of the observables $O_2, \ldots, O_6$ in Fig.~\ref{Plots of the observables O2, ..., O6}. The observables $O_7$, $O_8$, and $O_9$ 
are all zero if $\frac{g_2}{g_1}$ and $\frac{g_3}{g_1}$ are real (see Eqs.~(\ref{explicit expressions for the optimal observables})).

\begin{figure}[h]
\begin{minipage}[h]{0.47 \linewidth}
\center{\includegraphics[width=0.85\linewidth]{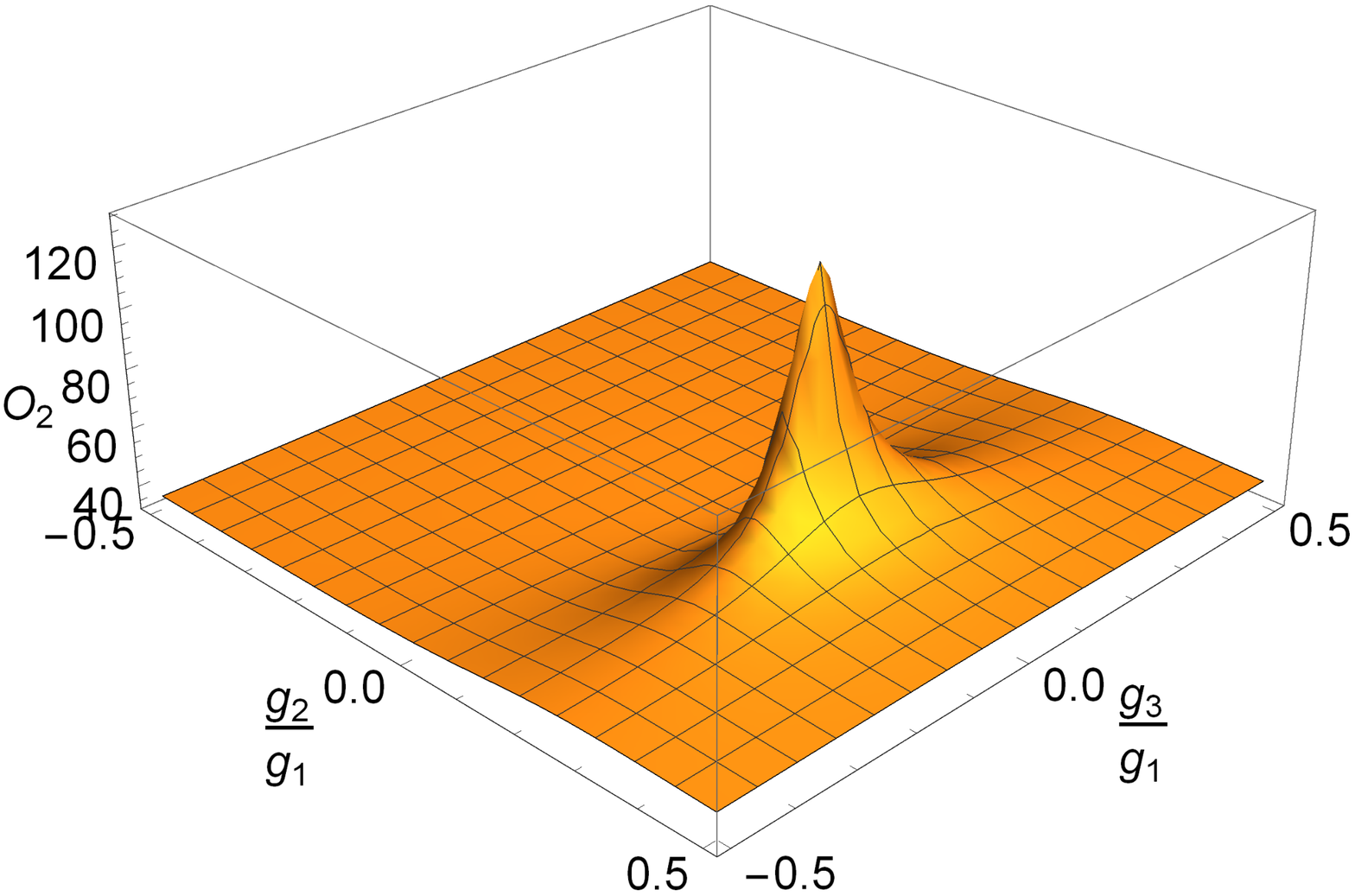}}
\end{minipage}
\hfill
\begin{minipage}[h]{0.47 \linewidth}
\center{\includegraphics[width=0.85\linewidth]{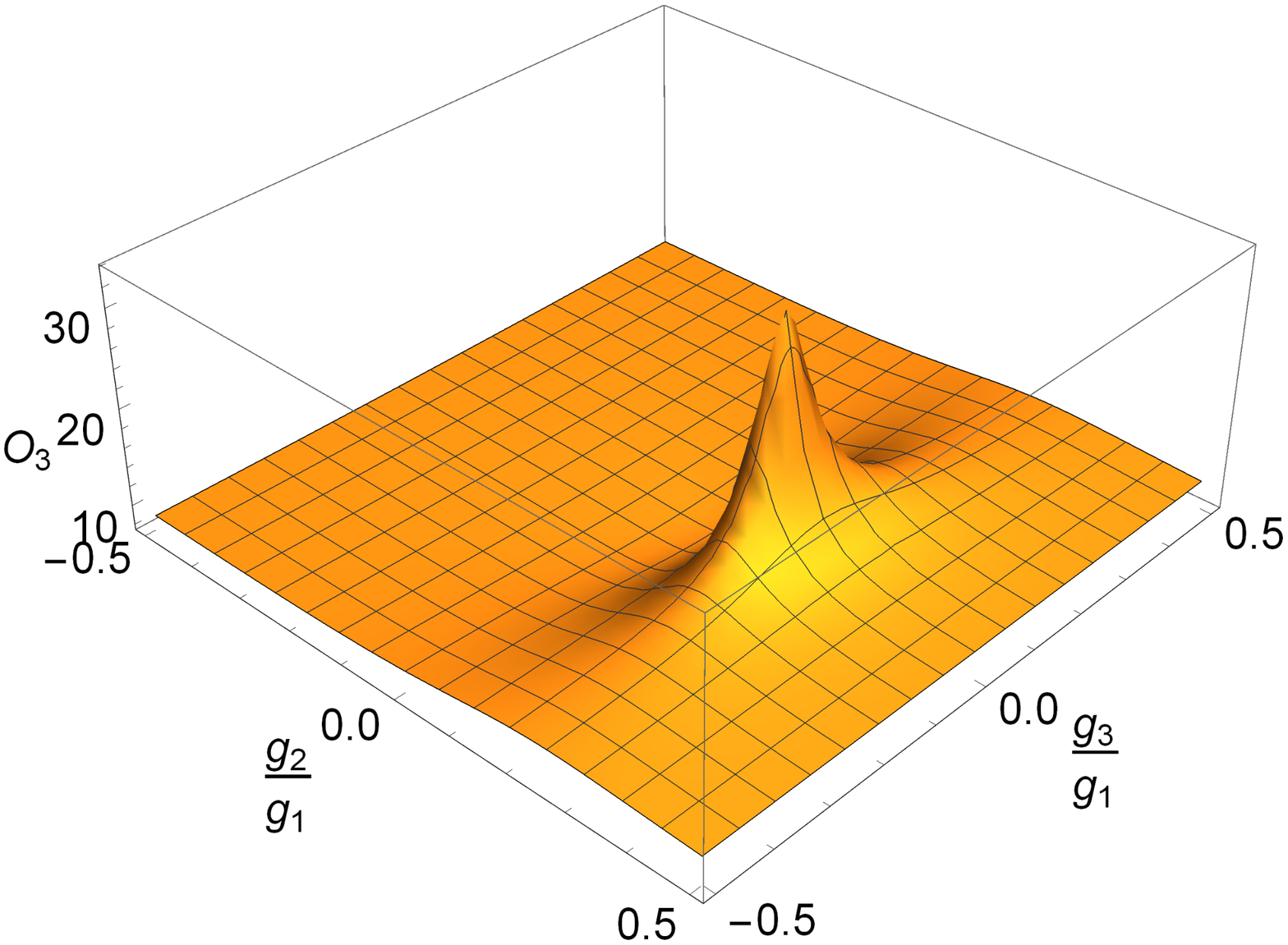}}
\end{minipage}
\vfill
\begin{minipage}[h]{0.47 \linewidth}
\center{\includegraphics[width=0.85\linewidth]{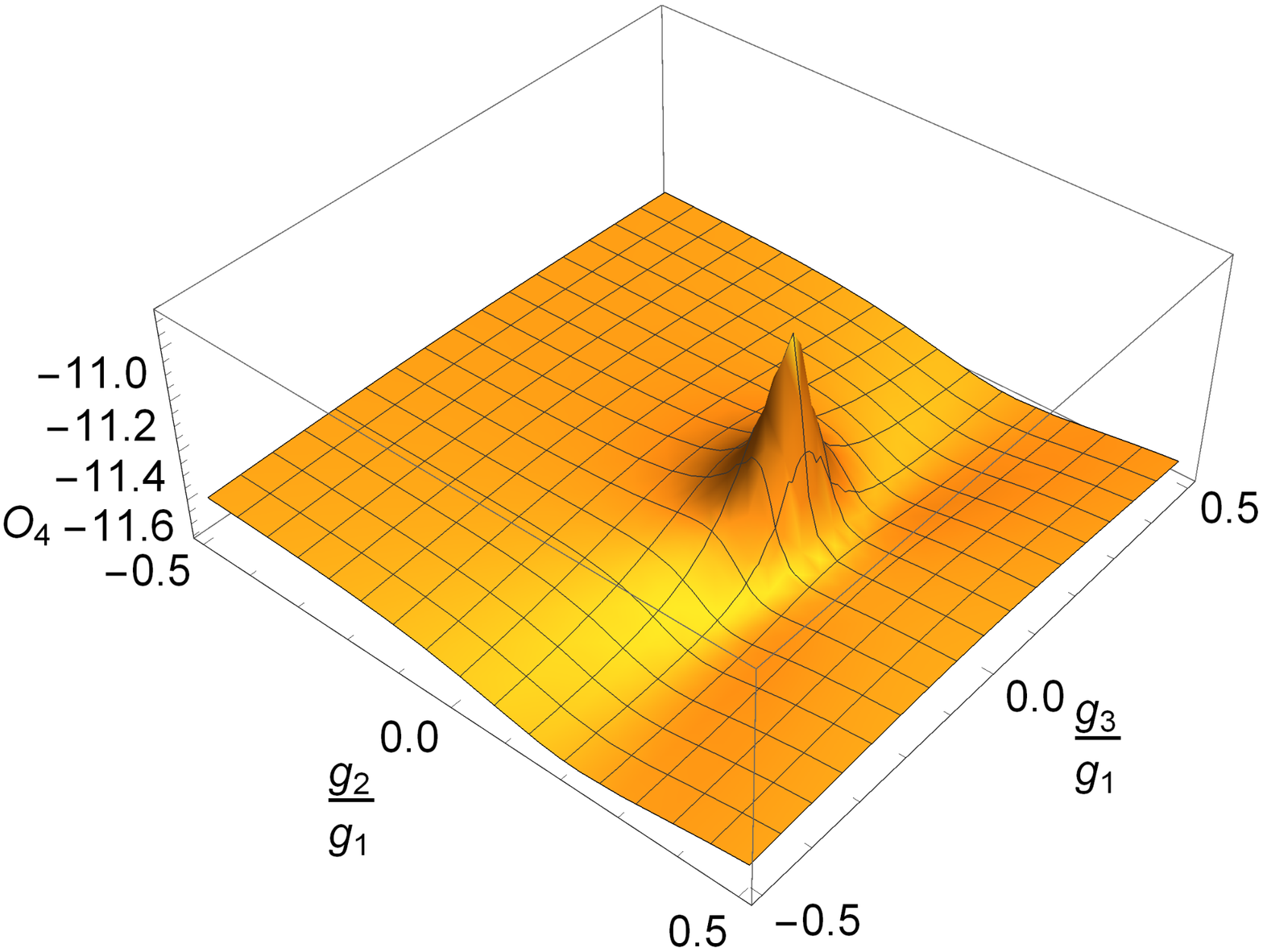}}
\end{minipage}
\vfill
\begin{minipage}[h]{0.47 \linewidth}
\center{\includegraphics[width=0.85\linewidth]{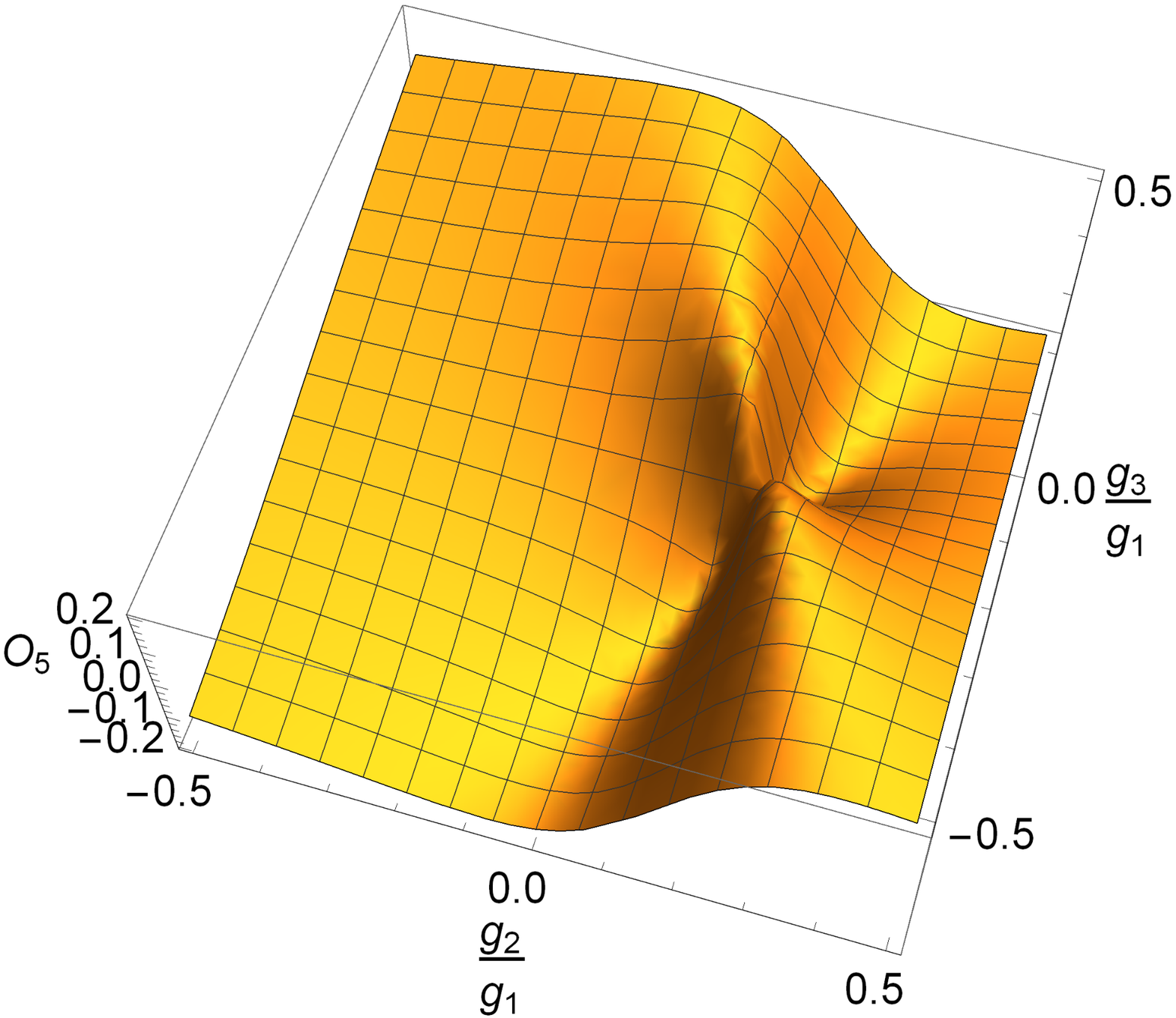}}
\end{minipage}
\hfill
\begin{minipage}[h]{0.47 \linewidth}
\center{\includegraphics[width=0.85\linewidth]{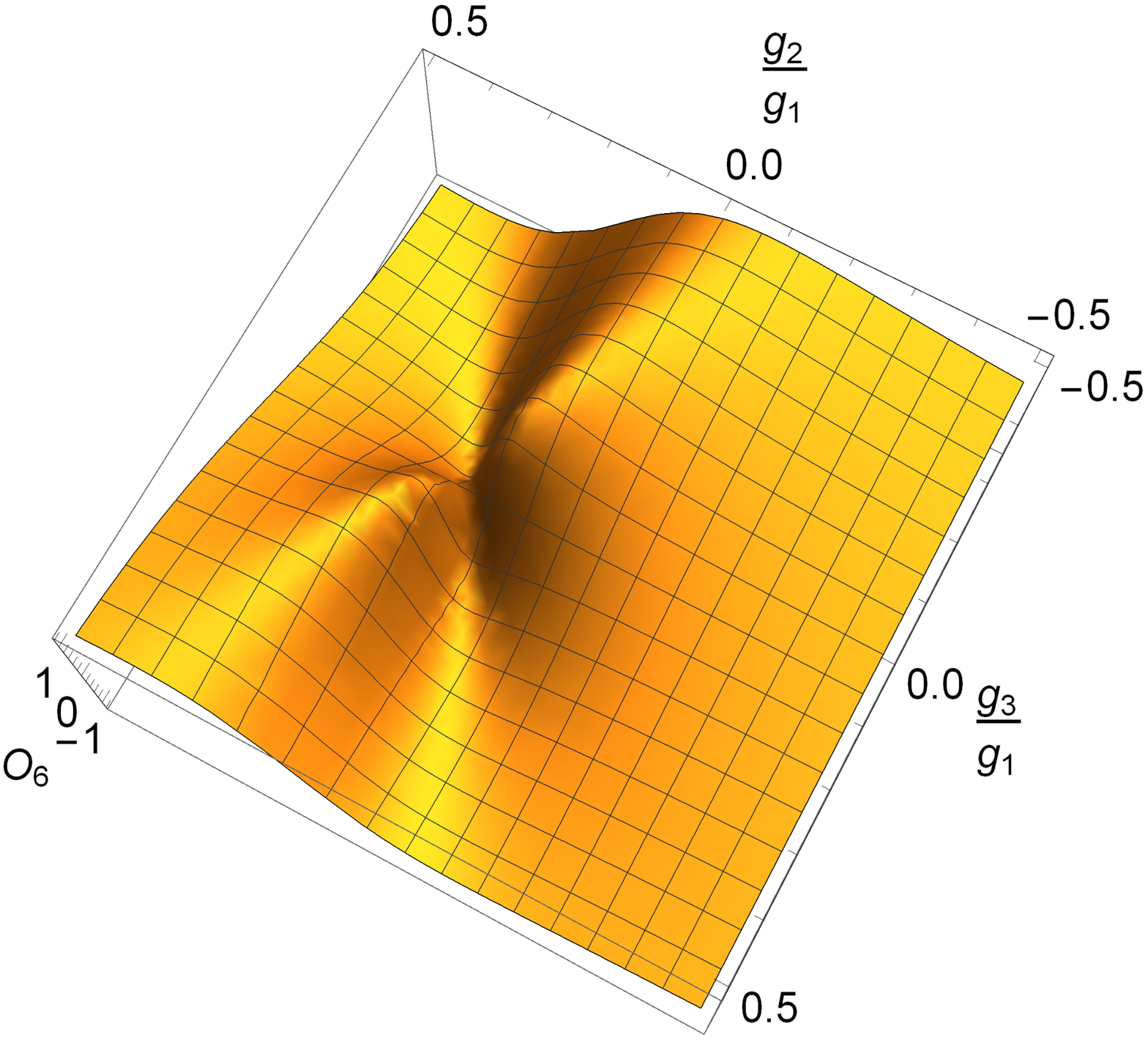}}
\end{minipage}
\caption{Plots of the observables $O_2, \ldots, O_6$ for $\frac{g_2}{g_1}, \frac{g_3}{g_1} \in [-0.5, 0.5]$.}
\label{Plots of the observables O2, ..., O6}
\end{figure}

The observables $O_2$, $O_3$, and $O_4$ are approximately proportional to each other. 
Moreover, we have analytically investigated the critical points of $O_2, \ldots, O_6$. Our calculation shows that each of $O_2$, $O_3$, and $O_4$ has a local maximum at $\frac{g_2}{g_1} \approx 0.17$ and $\frac{g_3}{g_1} = 0$. Both $O_5$ and $O_6$ have a saddle point at the same point $\frac{g_2}{g_1} \approx 0.17$, $\frac{g_3}{g_1} = 0$. The closer the values of $O_2, \ldots, O_6$ are to their values at the maximum or saddle points, 
the faster these observables change and the tighter constraints on $\frac{g_2}{g_1}$ and $\frac{g_3}{g_1}$ these observables provide.

If one calculates observables (\ref{optimal observables O2, O3, ..., O9}) for process (\ref{the Higgs-strahlung}) with $f = e$, $f_Z = e, \mu$ and $f_X = e, \mu, \tau, u, d, s, c, b$, then in the approximation $A_e = A_{\mu}$ the result still will be Eqs.~(\ref{explicit expressions for the optimal observables}). If you want to take into account the difference between $A_e$ and $A_{\mu}$ ($A_e = 0.1515$ and $A_{\mu} = 0.142$ \cite{the_2016_review_of_particle_physics_by_the_Particle_Data_Group}), you can edit Supplemental Material \cite{a_Mathematica_notebook}. Moreover, in an experiment events (\ref{the Higgs-strahlung}) cannot be measured with all the possible values of the kinematic variables $k$. To account for that, one can change the limits of integration for observables (\ref{optimal observables O2, O3, ..., O9}) in \cite{a_Mathematica_notebook}.

\section{Conclusions}
\label{Section: Conclusions}

We have derived the fully differential cross section of the Higgs-strahlung process $f \bar{f} \to Z \to Z X \to f_Z \bar{f}_Z f_X \bar{f}_X$, where $f$, $f_Z$, and $f_X$ are arbitrary fermions, $X$ is a particle with zero spin and arbitrary $ZZ$ and $ff$ couplings. This process with $f = e$, $X = h$, $f_Z = e, \mu, u, d, s, c, b$, and $f_h = \tau, b$ is going to be measured at the ILC. 

We have defined observables measurement of which gives the tightest constraints on the Higgs boson couplings $g_1$, $g_2$, $g_3$ to $Z$ bosons. These observables are called optimal. Their explicit dependences on $g_i$ are presented. The sensitivity of the observables to the couplings is analyzed. Therefore, after the optimal observables are measured at the ILC, one can constrain the Higgs boson couplings $g_i$. Using \cite{a_Mathematica_notebook}, one can account for experimental constraints on the kinematic variables or derive explicit expressions of optimal observables for other similar processes. 

\section*{Acknowledgments}
This research was partially supported by the National Academy of Sciences of Ukraine (project no. Ts-3/53-2018) 
and the Ministry of Education and Science of Ukraine (project no. 0117U004866).


\begin{thebibliography}{1}
\bibitem{a_Mathematica_notebook}
A Mathematica notebook where the optimal observables for the Higgs-strahlung are worked out is available from the authors upon request.
\end{thebibliography}
\end{document}